\documentclass{iopconfser}

\usepackage{amsmath, amssymb}
\usepackage{mathrsfs}
\usepackage[dvipdfmx]{graphicx}
\usepackage{bm}
\usepackage{braket}
\usepackage{multirow}
\usepackage[colorlinks=true, citecolor=blue]{hyperref}
\usepackage{floatrow,ulem,ascmac}

\newcommand{\diff}{\mathrm{d}}

\newcommand{\trace}{\mathrm{Tr}\,}
\newcommand{\imu}{\mathrm{i}}
\newcommand{\epn}{\mathrm{e}}

\newcommand{\ua}{\uparrow}
\newcommand{\da}{\downarrow}
\newcommand{\dg}{\dagger}
\newcommand{\la}{\langle}
\newcommand{\ra}{\rangle}
\newcommand{\al}{\alpha}

\newcommand{\sg}{\sigma}
\newcommand{\om}{\omega}
\newcommand{\gm}{\gamma}

\newcommand{\ep}{\varepsilon}

\newcommand{\mcal}[1]{\mathcal{#1}}

\usepackage{color}

\usepackage{CJKutf8}

\begin{document}

\title{
Weak coupling approach to magnetic and orbital susceptibilities for superconducting states in multiorbital electron-phonon coupled model
}

\author{Natsuki Okada$^{1}$, Tatsuya Miki$^{1}$ and Shintaro Hoshino$^{1}$}

\affil{$^1$Department of Physics, Saitama University, Saitama, Japan}

\email{n.okada.473@ms.saitama-u.ac.jp}

\begin{abstract}
Alkali-doped fullerides are molecular-based superconductors with multiple active orbitals.
In this paper, using the Eliashberg theory with the retardation effect of Jahn-Teller phonons,
we study the response of the spin-singlet superconducting state relevant to fulleride materials.
The spin Zeeman field is not active for the singlet pairing state, and the magnetic orbital field, which physically generates a circular electron motion inside the fullerene molecule, is also shown to be inactive.
On the other hand, the electric orbital (or quadrupolar) field, which corresponds to a uniaxial distortion, remains active across the superconducting phase transition.
This is understood by the orbital-symmetric structure of the Cooper pair, which is susceptible to the electric orbital field, while it is not the case for the magnetic orbital field which tends to create an antisymmetric part.
\end{abstract}

\noindent



\section{Introduction}

In molecular-based crystals, the relevant phonon contributions stem from local molecular vibrations, and the interaction effect inside each molecule plays a significant role. 
A basic model to capture this is the Holstein-Hubbard model where single-orbital electrons couple with spatially local Einstein phonons \cite{Holstein}. On the other hand, fulleride superconductors involve three degenerate molecular orbitals, $t_{1u}$, which are coupled to spatially local, anisotropic molecular vibrations (Jahn-Teller phonons) \cite{Gunnarsson97}. To understand fulleride superconductors, it is essential to analyze the Jahn-Teller-Hubbard model, incorporating multiple electron and phonon degrees of freedom \cite{Capone09,Nomura16}.

Because of the multiorbital nature, the fulleride superconductor has more degrees of freedom than the standard BCS superconductor. In this paper, we aim to study the effect of molecular orbital degrees of freedom on the superconducting state. More specifically, we apply the small but finite magnetic field and directional stress and observe the conjugate physical quantities such as spin and orbital moments.

As for the theoretical framework, in three-dimensional fullerides, intersite correlations remain relatively weak, making the dynamical mean-field theory (DMFT) a suitable method due to its ability to account for local correlations \cite{Georges96}. 
In this paper, we combine DMT with the Eliashberg theory \cite{Eliashberg60,Eliashberg63,Wada64,Bardeen64,Parks_book}, which is valid for the weak-coupling regime, to analyze the various spin and orbital susceptibilities in the multiorbital electron-phonon coupled model \cite{Han00,Han03,Nomura15,Kaga22}. 

As demonstrated in the following, the spin susceptibility is quickly suppressed below $T_c$ as in an ordinary single-orbital superconductor. 
A similar behavior is also observed for the magnetic orbital field.
On the other hand, for the perturbation with respect to electric orbital degrees of freedom (i.e. quadrupole) shows that their response is not influenced by the superconductivity although the single-particle spectrum has a gaped structure.
We interpret this result by focusing on the symmetric orbital structure of the Cooper pair, which is affected by symmetric perturbation such as electric orbital distortion, but not the case for the magnetic orbital field which tends to create the antisymmetric part.
We also analyze a simple multiorbital BCS Hamiltonian with no retardation effects, which also supports our conclusion.

This paper is organized as follows.
The next section describes the formalism of the Eliashberg equation.
Section 3 is devoted to the introduction of the susceptibilities and their method of evaluation.
Section 4 shows the detailed numerical results and their interpretation based on a simplified BCS model without retardation effects.
We summarize our paper in Section 5.

\section{Coupling to Jahn-Teller phonons}

\subsection{Model}

We consider a multiorbital electron model coupled with anisotropic vibrational modes of the fullerene molecules
\cite{Kaga22}.
The total Hamiltonian is given by $\mathscr{H} = \mathscr{H}_{0}  
+ \mathscr{H}_{\mathrm{ep}}$, where 
the non-interacting part $\mathscr{H}_{0}$ is given by
\begin{align}
    \mathscr{H}_{0} &= 
    \sum_{ij}\sum_{\gm\gm'\sg} 
    \left(  t_{ij}^{\gm\gm'} -\mu \delta_{ij} \delta_{\gm\gm'} 
    \right) c_{i\gm\sg}^\dg c_{j\gm'\sg}
    +
    \sum_{i}\sum_{\substack{\eta=0,1,3,\\4,6,8}} \omega_\eta a_{i\eta}^\dg a_{i\eta}, \label{eq:ham_0}
\end{align}
Each term corresponds to a kinetic energy of electrons and local phonons, respectively.
For the electron part, we have introduced site indices $i, j$, spin index $\sg=\ua, \da$, and an index for $t_{1u}$ molecular orbitals $\gamma=x, y, z$.
For the phonon part, we have taken into account the isotropic $A_g$ mode ($\eta = 0$) and anisotropic $H_g$ mode ($\eta=
1,3,4,6,8$), whose energies are represented as $\om_\eta$.
In this paper, we focus on the electron-phonon coupling and neglect the Coulomb interaction among electrons for simplicity.

The electron-phonon interaction parts $\mathscr{H}_{\mathrm{ep}}$ is given by
\begin{align}
  \mathscr{H}_{\mathrm{ep}}
  &= \sum_{i}\sum_{\substack{\eta=0,1,3,\\4,6,8}} \sum_{\gm\gm'\sg} g_\eta \phi_{i\eta} c_{i\gm\sg}^\dg  \hat{\lambda}_{\gm\gm'}^\eta c_{i\gm'\sg}
  = \sum_{i} \sum_{\substack{\eta=0,1,3,\\4,6,8}} g_\eta \phi_{i\eta} T_{i\eta}, \label{eq:ham_ep}
\end{align}
where $\phi_{i\eta} = a_{i\eta} + a_{i\eta}^\dg$ is a displacement operator of the fullerene molecules. 
In Eq.~\eqref{eq:ham_ep}, we have introduced the charge-orbital moment $T_{i\eta} = \sum_{\gm\gm'\sg} c_{i\gm\sg}^\dg \hat\lambda_{\gm\gm'}^\eta c_{i\gm'\sg}$ \cite{Iimura21}
with Gell-Mann matrices $\hat\lambda^\eta$, 
\begin{align}
  \hat{\lambda}^0 &= \sqrt{\frac{2}{3}}
  \begin{pmatrix}
    1 & 0 & 0 \\
    0 & 1 & 0 \\
    0 & 0 & 1 
  \end{pmatrix}
  , \hspace*{3mm}
  \hat{\lambda}^1 =
  \begin{pmatrix}
    0 & 1 & 0 \\
    1 & 0 & 0 \\
    0 & 0 & 0
  \end{pmatrix}
  , \hspace*{3mm}
  \hat{\lambda}^2 =
  \begin{pmatrix}
    0 & -\imu & 0 \\  
    \imu & 0 & 0 \\
    0 & 0 & 0 
  \end{pmatrix}
  , \hspace*{3mm}
  \hat{\lambda}^3 =
  \begin{pmatrix}
  1 & 0 & 0 \\  
  0 & -1 & 0 \\
  0 & 0 & 0
  \end{pmatrix}
  , \hspace*{3mm}
  \hat{\lambda}^4 = 
  \begin{pmatrix}
    0 & 0 & 1 \\
    0 & 0 & 0 \\
    1 & 0 & 0 
  \end{pmatrix}
  , 
  \nonumber\\
  \hat{\lambda}^5 &= 
  \begin{pmatrix}
    0 & 0 & -\imu \\
    0 & 0 & 0 \\
    \imu & 0 & 0
  \end{pmatrix}
  , \hspace*{3mm}
  \hat{\lambda}^6 = 
  \begin{pmatrix}
    0 & 0 & 0 \\
    0 & 0 & 1 \\
    0 & 1 & 0
  \end{pmatrix}
  , \hspace*{3mm}
  \hat{\lambda}^7 =
  \begin{pmatrix}
    0 & 0 & 0\\
    0 & 0 & -\imu \\
    0 & \imu & 0
  \end{pmatrix}
  , \hspace*{3mm}
  \hat{\lambda}^8 = \sqrt{\frac{1}{3}}
  \begin{pmatrix}
    1 & 0 & 0 \\
    0 & 1 & 0 \\
    0 & 0 & -2
  \end{pmatrix}
  . \label{GEll-mann}
\end{align}
%
%
The $\eta=0$ component describes the charge degrees of freedom (irreducible representation $A_g$ in point group $I_h$).
The anisotropic orbital or quadrupolar moments are characterized by $\eta= 1$ ($xy$ type), $\eta=6$ ($yz$ type), $\eta = 4$ ($zx$ type), $\eta=8$ ($3z^2-r^2$ type) , $\eta=3$ ($x^2-y^2$ type), which belong to the irreducible representation $H_g$.
While there are several vibration modes coupled to $t_{1u}$ electron orbital, we here consider the representative ones with the largest electron-phonon coupling \cite{Kaga22}.
We note that the absence of $\eta = 2$ ($z$ type),  $\eta = 5$ ($y$ type), and  $\eta = 7$ ($x$ type) components in the Hamiltonian, which physically describe orbital angular momentum (or dipole moments), can be confirmed by using the irreducible decomposition of the electronic degrees of freedom \cite{Kaga22}.
On the other hand, the antisymmetric $\eta=2,5,7$ matrices will be used for the external field part (see Sec.~3).

For later convenience, we introduce the Pauli matrices as follows:
\begin{align}
    \sg^0 = 
    \begin{pmatrix}
        1 & 0 \\
        0 & 1
    \end{pmatrix}, 
    \sg^1=
    \begin{pmatrix}
        0 & 1 \\
        1 & 0
    \end{pmatrix},  
    \sg^2 = 
    \begin{pmatrix}
        0 & -\imu \\
        \imu & 0
    \end{pmatrix}, 
    \sg^3 = 
    \begin{pmatrix}
        1 & 0 \\
        0 & -1
    \end{pmatrix}. 
\end{align}
The antisymmetric matrix is also defined by $\epsilon = \imu\sg^y$.

\subsection{Local Green's functions and self-energies}

We include the electron-phonon interaction effect by using the perturbation theory, which is valid in the Fermi liquid regime.
Furthermore, we assume that the self-energy, which reflects the interaction effect, is dominantly local in a three-dimensional fulleride superconductor.
Hence we apply the DMFT \cite{Georges96} combined with lowest-order perturbation theory.

We begin with the definition of the local Green's function for electrons.
We introduce the Nambu spinors by 
$\bm \Psi_\gm = (c_{\gm\ua}, c_{\gm\da}^\dg)^T$ and its conjugate by $\bm \Psi_\gm^\dg = (c_{\gm\ua}^\dg, c_{\gm\da})$, 
and then the Green's function is given by
\begin{align}
   \check{G}_{\gm\gm'}(\tau) 
   &= -\la \mathcal{T}\,\bm \Psi_\gm(\tau)\bm \Psi_{\gm'}^\dg(0) \ra, 
\end{align}
where the site index is omitted by assuming the translational invariance in solids.
The $3\times 3$ matrix in orbital space is denoted by a hat symbol ( $\hat{}$ ), and the $6\times 6$ matrix in Nambu-orbital space is denoted by a check symbol ( $\check{}$ ).
$A(\tau) = \epn^{\tau \mathscr H}A \epn^{-\tau \mathscr H}$ is the Heisenberg representation with imaginary time, and $\mathcal T$ represents the imaginary-time ordering operator.

The local Green's function is obtained once the local self-energy matrix $\check \Sigma(\imu\omega_n)$ is given.
The explicit relation is written down as
\begin{align}
    \check{G}(\imu\omega_n) 
    &= 
    \int_{-D}^{D} \diff\ep 
    \rho(\ep)
    \left[
    \imu\omega_n \check{1} - (\ep - \mu)\check{\tau}_3 -\check H_{\rm ext} - \check{\Sigma}(\imu\omega_n)
    \right]^{-1}, 
\end{align}
where $\check{\tau}_3 = \mathrm{diag}(\hat{1}, -\hat{1})$.
$\omega_n = (2n+1)\pi T$ ($n\in\mathbb{Z}$) is the fermionic Matsubara frequency, 
$\mu$ is the chemical potential, and $\check H_{\rm ext}$ is the external field part defined in the next section.
The integral with respect to $\ep$ corresponds to the wave-vector integral needed for obtaining the local Green's function.
For simplicity, the density of states is chosen to be featureless semi-circular shape:
\begin{align}
    \rho(\ep) = \frac{2}{\pi D^2}\sqrt{D^2 - \ep^2}, 
\end{align}
where the bandwidth is given by $2D$.
The actual numerical calculation for $\ep$ integral is performed using the Gauss-Legendre method.

On the other hand,
the phonon Green's function is defined by the displacement operator $\phi_\eta$
\begin{align}
    D_{\eta\eta'}(\tau) &= -\la\mathcal{T}\, \phi_\eta(\tau) \phi_{\eta'}(0)\ra, 
\end{align}
which is $6\times 6$ matrix 
with respect to $\eta, \eta'$. 

Once the local phonon self-energy matrix $\Pi$ is given, the phonon Green's function is obtained in the frequency domain as 
\begin{align}
    D^{-1}(\imu\nu_m) &= D_{0}^{-1}(\imu\nu_m) - \Pi (\imu\nu_m),
    \\
    D_{0, \eta\eta'}(\imu\nu_m)
    &=
    \frac{2\omega_\eta}{(\imu\nu_m)^2 - \omega_\eta^2}\delta_{\eta\eta'}, 
\end{align}
where $\nu_m = 2m\pi T$ ($m\in\mathbb{Z}$) is the bosonic Matsubara frequency.
$D_0$ is the non-interacting Green's function.
Since we focus on the response of the electrons against external perturbations in this paper, 
we do not consider the external field part for phonons. 

\subsection{Determination of self-energy: Eliashberg equation}

In general, the self-energies introduced in the previous subsection are given by a functional of Green's functions.
In the second-order perturbation theory with respect to the electron phonon coupling $g_\eta$, we obtain the following form of the self-energies for electrons and phonons \cite{Kaga22,Ishida24}:
\begin{align}
    \check \Sigma_{\mathrm{ep}} (\tau) &=
    - 
    \sum_{\eta, \eta'}
    g_\eta g_{\eta'}D_{\eta\eta'} (\tau)
    \Bigl[
        \check{\lambda}^{\eta} 
        \check{G}(\tau)
        \check{\lambda}^{\eta'}
    \Bigr], 
    \\
    \Pi_{\eta\eta'}(\tau) &=
    \frac 1 2
    g_{\eta}g_{\eta'}  
    \,
    \trace
    \Bigl[
        \check{G} (- \tau) \check{\lambda}^{\eta'}
        \check{G} (\tau ) \check{\lambda}^{\eta}
    \Bigr], 
\end{align}
where we have defined the Gell-Mann matrix extended to Nambu space as
\begin{align}
    \check{\lambda}^\eta &=
    \begin{pmatrix}
        \hat{\lambda}^\eta & 0 \\
        0 & -\hat{\lambda}^{\eta T}
    \end{pmatrix}
    . 
\end{align}
The equation system is thus closed.
The actual calculation is performed in an iterative way until the converged solution is obtained.


\section{Susceptibility}

\subsection{General remarks}

In this subsection, we explain the calculation method for the various susceptibilities.
We denote the physical quantity by $\mathscr O$
and its conjugate field is given by $h$.
Then the susceptibility is defined by
\begin{align}
    \chi
    &=
    \lim_{h\to 0} \frac{\partial \la \mathscr O \ra}{\partial h}
    =\lim_{h\to 0}
    \frac{\la \mathscr{O}
    \ra
    |_{h} - \la \mathscr{O}
    \ra|_{h=0}}{h}. 
\end{align}
The corresponding external perturbation Hamiltonian is given by
\begin{align}
    \mathscr H_{\rm ext} &= -h \mathscr O = \sum_i 
    \bm \Psi_i^\dg \check H_{\rm ext} \bm \Psi_i. 
\end{align}
In the actual calculation, we set a small but finite $h$ and perform self-consistent calculation, and evaluate the conjugate physical quantity.
We then calculate the quantity at $h=0$ and obtain the susceptibility $(\la \mathscr{O}
    \ra
    |_{h} - \la \mathscr{O}
    \ra|_{h=0})/h$.
Below, we consider both the diagonal and offdiagonal quantities.

\subsection{Diagonal susceptibility: Spin response}

For the magnetic moment, we consider the following quantities:
\begin{align}
    S^z
    &=
    \sum_{\gm}\sum_{\sg\sg'}
    c_{\gm\sg}^\dg \sg_{\sg\sg'}^3 c_{\gm\sg'}.  \label{Sz}
\end{align}
The external field part, which is identical to the spin Zeeman effect, is given by
\begin{align}
    \check H_{\rm ext}
    &=
    -h
    \begin{pmatrix}
        \hat 1 & \hat{0} \\
        \hat{0} &  \hat 1
    \end{pmatrix}. 
\end{align}

\subsection{Diagonal susceptibility: Orbital response}

We also consider other diagonal quantities related to the orbital degrees of freedom:
\begin{align}
    L^z
    &=
    \sum_{\gm\gm'}\sum_{\sg}
     c_{\gm\sg}^\dg \hat\lambda_{\gm\gm'}^2 c_{\gm'\sg},  \label{Lz}
    \\
    Q^{x^2-y^2}
    &=
    \sum_{\gm\gm'}\sum_{\sg}
   c_{\gm\sg}^\dg  \hat\lambda_{\gm\gm'}^3 c_{\gm'\sg}  \label{Qx^2-y^2}
    , 
\end{align}
each of which corresponds to the magnetic orbital, and electric orbital (or quadrupolar) moments, respectively.

The corresponding external field matrix element for the above quantities is written as
\begin{align}
    \check H_{\rm ext}
    &=
    -h
    \begin{pmatrix}
        \hat\lambda^\eta & \hat{0} \\
        \hat{0} &  -\hat\lambda^{\eta\,T}
    \end{pmatrix}. 
\end{align}

\subsection{Offdiagonal susceptibility: Spin-singlet pairing}
In fulleride superconductors, the orbital-symmetric spin-singlet pair is known to be realized \cite{Han03,Nomura15,Kaga22}.
The order parameter is given by
\begin{align}
    p^{s-\mathrm{wave}}
    &=
    \sum_{\gm}
    c_{\gm\ua}^\dg c_{\gm\da}^\dg + \mathrm{H.\,c.}\,. 
    \label{singlet_pair}
\end{align}
The corresponding matrix element for the external field part is 
\begin{align}
    \check H_{\rm ext}
    &=
    -h
    \begin{pmatrix}
         \hat{0} & \hat{1}\\
         -\hat{1} & \hat{0}
    \end{pmatrix}. 
\end{align}


\section{Results }
In this section, we solve the Eliashberg equations within the framework of the weak coupling limit and present the resulting susceptibility. The bandwidth is set to $2D=1.0\,\mathrm{eV}$, the chemical potential to $\mu=0\,\mathrm{eV}$, and the test field to $h=0.001\,\mathrm{eV}$. 
Because of the small but finite test field ($h\neq 0$), the susceptibilities are not divergent in numerical calculations.
For simplicity, the phonon energy is taken to be $\omega_\eta=0.25\,\mathrm{eV}$, regardless of the type of vibrational mode. The dimensionless electron-phonon coupling constants are set as $\lambda_0=0.1$ and $\lambda_{1,3,4,6,8}=\lambda_1=0.03$.

\subsection{Numerical simulation}

\begin{figure*}[tb]
    \centering
    \includegraphics[width=100mm]{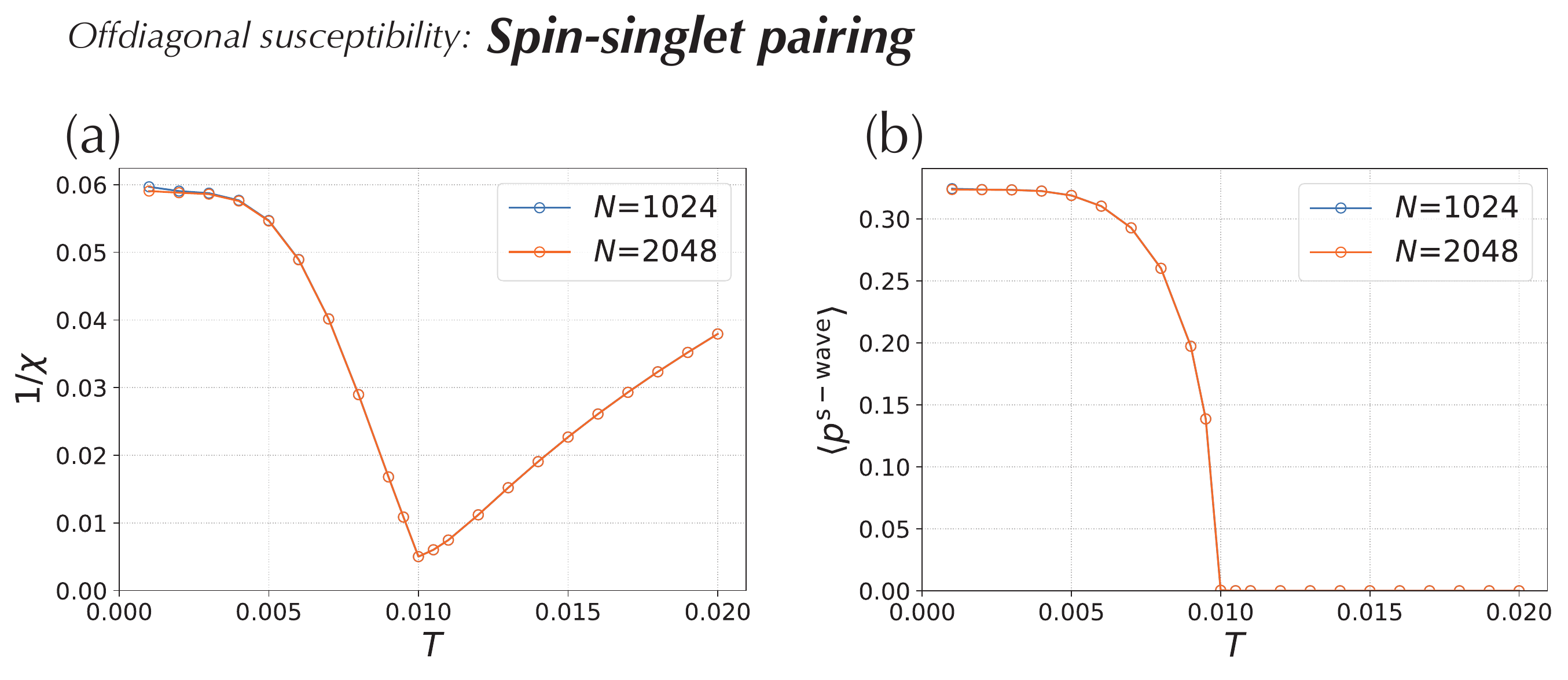}
    \caption{(a) Temperature dependence of the inverse susceptibility for spin-singlet pairing. (b) Temperature dependence of the spin-singlet pair amplitude in the absence of an external field. 
    Calculations in (a) and (b) were conducted for different numbers of Matsubara frequencies $N=1024$ and $N = 2048$
    .}
    \label{fig:T_singlet}
\end{figure*}

To elucidate the impact of molecular orbital degrees of freedom on the superconducting state, we first examine the temperature range in which the superconducting state emerges.
Fig.\,\ref{fig:T_singlet}(a) shows the temperature dependence of the inverse susceptibility of the spin-singlet pairing.
At the temperature $T = 0.01$, the inverse susceptibility reaches a minimum value and then increases gradually as the temperature decreases further. 
In the figures, we show the two sets of data for different numbers of Matsubara frequencies ($N=1024$ and $N=2048$), to validate our numerical results except in the low-temperature limit.

Fig.\,\ref{fig:T_singlet}(b) shows the temperature dependence of the spin-singlet pair amplitude, i.e., the order parameter, in the absence of an external field. At high temperatures, the order parameter is zero, while it becomes finite at the transition point $T = 0.01$ (denoted as $T_c$), and reaches a constant value at lower temperatures. This behavior indeed represents a second-order phase transition characteristic of superconductivity, confirming the emergence of the superconducting phase at temperatures below $T_c$.

\begin{figure}[tb]
    \centering
    \includegraphics[width=155mm]{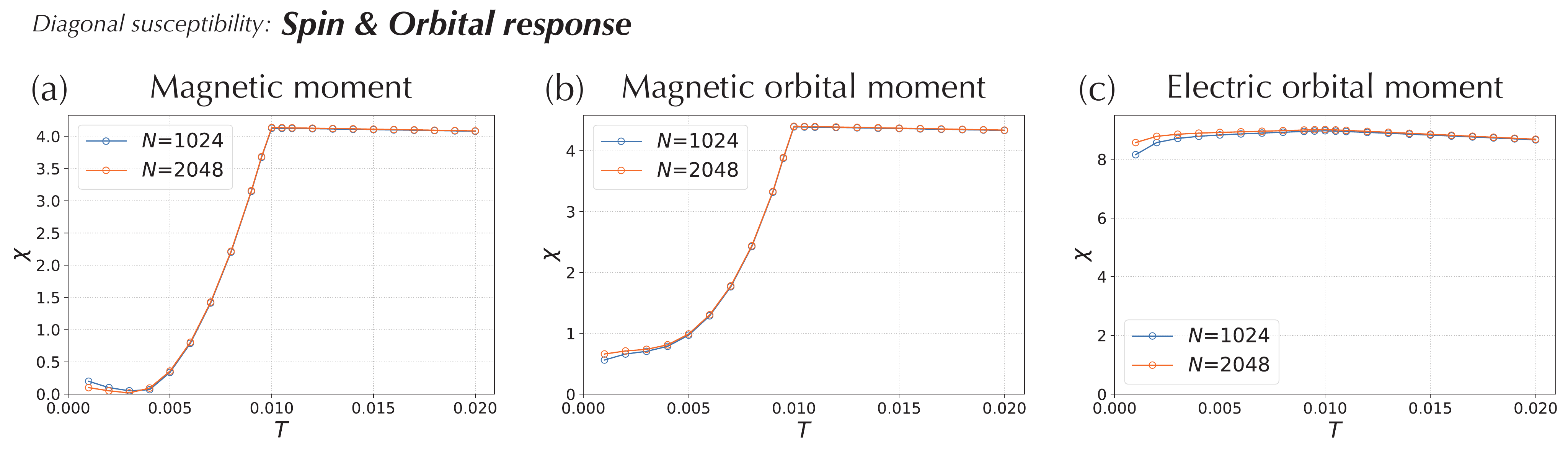}
    \caption{
    Temperature dependence of the susceptibilities for (a) magnetic moments, (b) magnetic orbital moments, and (c) electric orbital moments. The calculations were conducted using the same parameters as in Fig.\,\ref{fig:T_singlet}.
    }
    \label{fig:T_without_singlet}
\end{figure}
Next, by applying conjugate external fields to the magnetic orbital, magnetic orbital moment, and electric orbital moment, we investigate the behavior of these susceptibilities in the superconducting phase. First, the temperature dependence of the spin susceptibility is shown in Fig.\,\ref{fig:T_without_singlet}(a). At temperatures above $T_c$, the spin susceptibility remains constant, while below $T_c$, it decreases and eventually reaches zero. 
This behavior is similar to the temperature dependence of the spin susceptibility in conventional single-orbital systems, where the magnetization below the superconducting transition temperature is proportional to the Yoshida function.

Next, we discuss the behavior of the orbital moments in the superconducting phase. 
The temperature dependence of the magnetic orbital susceptibility is shown in Fig.\,\ref{fig:T_without_singlet}(b). Similar to the case of the magnetic moment in Fig.\,\ref{fig:T_without_singlet}(a), the susceptibility remains constant at temperatures above $T_c$, and decreases as the temperature drops below $T_c$. However, in contrast to the magnetic moment case, the orbital susceptibility remains finite at low temperatures. 
We will further discuss this point in Sec. 4.2 by comparing it with BCS theory without the retardation effect of electron-phonon coupling.

%
%
%
Finally, the temperature dependence of the electric orbital susceptibility is shown in Fig.\,\ref{fig:T_without_singlet}(c). In the case of the electric orbital moment, the temperature dependence of the susceptibility remains constant even 
inside
the superconducting phase. This behavior is qualitatively different from that of the magnetic moment and magnetic orbital moment. The unique behavior of the electric orbital moment is attributed to the symmetric structure of the matrix element in Eq.~\eqref{GEll-mann}.
This will be analytically explained in the next subsection (Sec. 4.2).

\subsection{BCS theoretical analysis of spin and orbital susceptibilities}

For comparison, let us consider the simple multiorbital BCS model for the spin and magnetic orbital susceptibilities.
We begin with the BCS model Hamiltonian given by
\begin{align}
    \mathscr{H}
    &=
    %
    %
    \sum_{\bm{k}\gamma\sg}\xi_{\bm{k}} c_{\bm{k}\gamma\sg}^\dg c_{\bm{k}\gm\sg}
    + \sum_{\bm{k}\gm}\Delta(c_{\bm{k}\gm\ua}^\dg c_{-\bm{k},\gm\da}^\dg + \mathrm{H.c.}), 
\end{align}
where we have assumed that the pair potential $\Delta$ is real-valued.
The structure of the Cooper pair is interpreted as $s$-wave, orbital-symmetric, and spin-singlet (spin-antisymmetric).
We add the external perturbation by
\begin{align}
\mathscr H_{\rm ext}   = - h_{\eta\mu} \sum_{\bm{k}}
    \sum_{\gamma\gamma'}
    \sum_{\sg\sg'}
    c_{\bm{k}\gm\sg}^\dg \lambda_{\gm\gm'}^\eta \sg_{\sg\sg'}^\mu c_{\bm{k}\gamma'\sg'}.     
\end{align}
We choose 
$(\eta,\mu)=(0,3)$ for the spin Zeeman field, 
$(\eta,\mu)=(2,0)$ for the magnetic orbital field, and 
$(\eta,\mu)=(3,0)$ for the electric orbital  field.
The susceptibility is calculated by the linear response theory 
\cite{Fetter03}
as
\begin{align}
    \chi_{\eta \mu}
    &\equiv
    \sum_{\bm{k}\bm{k}'} 
    \sum_{\gm_1\gm_1'} \sum_{\gm_2\gm_2'}
    \sum_{\sg_1\sg_1'} \sum_{\sg_2\sg_2'}
    \int_0^\beta \diff\tau
    \big\la 
    \mcal T
    c_{\bm{k}\gm_1\sg_1}^\dg(\tau) \lambda_{\gm_1\gm_1'}^\eta \sg^\mu_{\sg_1\sg_1'}c_{\bm{k}\gm_1'\sg_1'} (\tau)
    c_{\bm{k}'\gm_2\sg_2}^\dg \lambda_{\gm_2\gm_2'}^\eta \sg^\mu_{\sg_2\sg_2'}c_{\bm{k}'\gm_2'\sg_2'}
    \big\ra_{0,{\rm conn}} . 
\end{align}
The quantum statistical average $\la \cdots \ra_{0,{\rm conn}}$ is taken by the Hamiltonian without external field, where only the connected diagram contribution is considered.
We use the normal and anomalous Green's function of the form
\begin{align}
    G_{\bm k,\gm\sg,\gm'\sg'}(\tau) &= - \la \mathcal T c_{\bm k \gm \sg}(\tau) c^\dg_{\bm k\gm'\sg'} \ra 
    = G_{\bm k}(\tau) \delta_{\gm\gm'} \delta_{\sg\sg'},
    \\
    F_{\bm k,\gm\sg,\gm'\sg'}(\tau) &= - \la \mathcal T c_{\bm k\gm \sg}(\tau) c_{-\bm k,\gm'\sg'} \ra 
    = F_{\bm k}(\tau) \delta_{\gm\gm'} \epsilon_{\sg\sg'},
    \\
    F^\dg_{\bm k,\gm\sg,\gm'\sg'}(\tau) &= - \la \mathcal T c^\dg_{-\bm k,\gm \sg}(\tau) c^\dg_{\bm k\gm'\sg'} \ra 
    = F^*_{\bm k,\gm'\sg',\gm\sg}(\tau)
    = - F_{\bm k}(\tau) \delta_{\gm\gm'} \epsilon_{\sg\sg'}. 
\end{align}
All of these Green's functions are assumed to be real-valued in imaginary-time representation.
Then we obtain
\begin{align}
    \chi_{\eta \mu} &= - \sum_{\bm k} \int_0^\beta
    \big[ G_{\bm k}(\tau) G_{\bm k}(-\tau) 
    {\rm tr\,}(\hat \lambda^\eta)^2 \ 
    {\rm tr\,}(\sg^\mu)^2 
    - F_{\bm k}(\tau)^2 
    {\rm tr\,}(\hat \lambda^\eta \hat \lambda^{\eta {\rm T}} ) \ 
    {\rm tr\,}(\sg^\mu \epsilon \sg^\mu \epsilon)    \big]
    \nonumber\\
    &=- 4 T\sum_{\bm{k}n}
    \big[
    G_{\bm{k}}(\imu\omega_n) G_{\bm{k}}(\imu\omega_n)
    - s_\eta s_\mu 
    F_{\bm{k}}(\imu\omega_n) F_{\bm{k}}(-\imu\omega_n)
    \big], 
\end{align}
where we have performed the Fourier transformation.
The signatures are defined by
\begin{align}
    s_\eta &=
    {\rm tr\,}(\hat \lambda^\eta \hat \lambda^{\eta {\rm T}} ) \, \big/  \,  {\rm tr\,}(\hat \lambda^\eta)^2, 
    \\
    s_\mu &=
    {\rm tr\,}(\sg^\mu \, \epsilon \, \sg^\mu \, \epsilon) \, \big/ \,
    {\rm tr\,}(\sg^\mu)^2. 
\end{align}
The sign in front of the anomalous function is an important factor in determining the susceptibility behavior:
(i) For the spin field, $(\eta,\mu)=(0,3)$, we have $s_\eta = 1$ and $s_\mu = 1$; 
(ii) For the magnetic orbital field, $(\eta,\mu)=(2,0)$, we have $s_\eta = -1$ and $s_\mu = -1$;
(iii) For the electric orbital field, $(\eta,\mu)=(3,0)$, we have $s_\eta = 1$ and $s_\mu = -1$.
Hence the cases (i) and (ii) show the same response with $s_\eta s_\mu = 1$, while the case (iii) is different from the others due to the reversed sign $s_\eta s_\mu = - 1$.
The origin of the sign is understood by focusing on the symmetric or antisymmetric structure of each matrix.


Let us evaluate the susceptibility further.
The explicit forms of the Green's functions are given by
\begin{align}
    G_{\bm{k}}(\imu\omega_n)
    &=
    \frac{\imu\omega_n + \xi_{\bm{k}}}{(\imu\omega_n)^2 - (\xi_{\bm{k}}^2+\Delta^2)}
    = \sum_{\al=\pm} \frac{a_{\bm{k}\al}}{\imu\omega_n - \al E_{\bm{k}}},
    \\
    F_{\bm{k}}(\imu\omega_n)
    &=
    \frac{\Delta}{(\imu\omega_n)^2 - (\xi_{\bm{k}}^2+\Delta^2)}
    = \sum_{\al=\pm} \frac{b_{\bm{k}\al}}{\imu\omega_n - \al E_{\bm{k}}}, 
\end{align}
where the single particle excitation energy for the Bogoliubov quasiparticles is given by $E_{\bm k} = \sqrt{\xi_{\bm{k}}^2+\Delta^2} >0$.
Utilizing partial fraction decomposition to derive the coefficients results in
    \begin{align}
        a_{\bm{k}\pm} &= \frac{1}{2}\left(1 \pm \frac{\xi_{\bm{k}}}{E_{\bm{k}}}\right), 
        \\
        b_{\bm{k}\pm} &= \pm \frac{\Delta}{2E_{\bm{k}}}. 
    \end{align}
Therefore, at zero temperature limit, we have
\begin{align}
    \chi_{\eta \mu}
    &=
    - 4\sum_{\bm{k}}\sum_{\al\al'} 
    (a_{\bm{k}\al} a_{\bm{k}\al'} - s_\eta s_\mu b_{\bm{k}\al} b_{\bm{k}\al'} )
    T\sum_n \frac{1}{(\imu\omega_n - \al E_{\bm{k}})(\imu\omega_n - \al' E_{\bm{k}})}
    \nonumber\\
    &\xrightarrow{T\rightarrow0}
    \ \ 
    4\sum_{\bm{k}}
    \frac{1}{E_{\bm{k}}}
    \left[
    a_{\bm{k}+} a_{\bm{k}-} 
    - s_\eta s_\mu
    b_{\bm{k}+} b_{\bm{k}-} )
    \right]
    \nonumber\\
    &=2(1-s_\eta s_\mu)
    \sum_{\bm{k}}
    \frac{\Delta^2}{2E_{\bm{k}}^3}
    \nonumber\\
    &\simeq 2(1-s_\eta s_\mu)
    \frac{D(0)}{2} \int_{-D}^{D} \diff\ep \frac{\Delta^2}{(\ep^2 + \Delta^2)^{3/2}}
    \nonumber\\
    &\simeq 2(1-s_\eta s_\mu)D(0), 
\end{align}
where $D(0)$ is the density of states at the Fermi level.
Namely, the susceptibility remains if $s_\eta s_\mu = -1$.
This result is qualitatively consistent with the numerical calculation of the Eliashberg equation.
Quantitatively, since the magnetic orbital susceptibility is not fully suppressed in the Eliashberg approach as shown in Fig.\,\ref{fig:T_without_singlet}(b), the retardation effect is more effective for the magnetic orbital moment than the spin moment. 


\section{Summary}

We constructed a multiorbital Eliashberg equation based on the Jahn-Teller Hubbard model and analyzed fulleride superconductors. Focusing on the 
effect of electron-phonon coupling, 
we examined the temperature dependence of the susceptibilities for spin-singlet pairing, magnetic moment, magnetic orbital moment, and electric orbital moment. Our results reveal that the magnetic orbital field suppresses the superconducting susceptibility, while the electric orbital susceptibility remains unaffected by the superconducting phase transition. The difference in these behaviors arises from the distinct orbital structures of the electric orbital field and the magnetic orbital field.

\section*{Acknowledgement}
N.O. was supported by Nisshin Seito Public Trust Scholarship Fund and Takafumi Horikawa Educational Foundation.
This work was supported by KAKENHI Grants No. 21K03459, No. 23KJ0298, and No. 24K00578.

\end{document}